\documentclass[aps,pra,twocolumn,showpacs,floatfix]{revtex4}
\usepackage{amsmath}
\usepackage{amssymb}
\usepackage{dsfont}

\begin{document}

\title{P\'olya number of continuous-time quantum walks}

\author{Z. Dar\'azs}
\email{zdarazs@optics.szfki.kfki.hu}
\author{T. Kiss}
\affiliation{Research Institute for Solid State Physics and Optics, Hungarian Academy of Sciences, P.O. Box 49, 1525 Budapest, Hungary}


\date{\today}

\begin{abstract}
We propose a definition for the P\'olya number of continuous-time quantum walks to characterize their recurrence properties. 
The definition involves a series of measurements on the system, each carried out on a different member from an ensemble in order to minimize the disturbance caused by it. We examine various graphs, including the ring, the line, higher dimensional integer lattices and a number of other graphs and calculate their P\'olya number. For the timing of the measurements a Poisson process as well as regular timing are discussed. We find that the speed of decay for the probability at the origin is the key for recurrence.
\end{abstract}

\maketitle

\section{Introduction}

Random walks are an important part of physics and statistical sciences \cite {revesz}. At the begining of the past century, George P\'olya considered the question, what is the probability that a walker ever returns to the starting point \cite{polyagy}. 
This probability is called  the P\'olya number. He found that in a one or two dimensional integer lattice the P\'olya number is one, {\it i. e.} the walker returns to the starting point with certainty, the walk is \textit{recurrent}. In higher dimensions, there is a finite probability that the walker never returns to the origin, in this case the walk is \textit{transient}. P\'olya's recurrence theorem is important in classical physics, for example in reaction-diffusion systems \cite{diffusion}.  

The concept of the quantum walk was introduced by Aharonov, Davidovich and Zagury \cite{aharonov}. The unitary time evolution of the quantum random walk has two main types: the discrete time evolution \cite{meyerwatrous}, when an extra coin degree of freedom is introduced and an associated coin operator assists in the evolution of the system at discrete time intervals, and the continuous-time case \cite{farhigut}, when the walker can change its position at any time. Classically, there is a straightforward embedding which relates the two types of walks to each other \cite{revesz}. The situation for the quantum walks is different, because the Hilbert space of discrete time quantum random walk contains an extra coin space \cite{ctqw-dtqw}.

Continuous-time quantum walks have recently been attracting interest, because they provide a tool to design efficient quantum algorithms \cite{algorithms}. In general, quantum walk can be regarded as a universal computational primitive, with any quantum computation encoded in some graph \cite{universal}. On the other hand, the physical relevance of continuous-time quantum walk stems from its relation to transport processes \cite{transport}.

Spreading of quantum walks is faster in general compared to the classical equivalent due to its wave nature \cite{fastspreading}, however interference can significantly modify this picture \cite{slowspreading}. The spreading properties of a quantum walk are related to the probability of staying at the localized starting position \cite{xu}, indicating, for example, the non-classical university class for a continuous-time quantum walk on a line with long-range steps \cite{long-range}. The time dependent  probability at a position is related to observation of the system if one additionally defines a measurement process. Our aim in this paper is to suggest such a process and define P\'olya number for this case and discuss how it is related to the asymptotic evolution.

Recurrence properties of the discrete time quantum walk can be characterized by the generalized P\'olya number\cite{dkpolyapra,dkpolyaprl}. In contrast to the classical case, the walk can be transient in two dimensions as well as recurrent in arbitrary higher dimensions, depending on the coin operator and the initial state. The definition does not translate directly to the continuous-time case since it contains measurements and, unlike in the discrete time case, there is no natural timing for detection.
Varbanov, Krovi and Brun encountered a similar problem in the definition of the hitting time for the continuous-time quantum walk \cite{varbanov}. They suggested to use a Poisson process for the timing, with $\lambda$ measurement rate. It means that in a small $\delta t$ period we measure in the system with $\varepsilon$ probability, and let the system evolve unitarity with $1-\varepsilon$ probability. 
 
In this paper, we extend our definition of the P\'olya number \cite{dkpolyapra,dkpolyaprl} to continuous-time quantum walks. The observation of recurrence includes measurements at different time instants, which unavoidably disturb unitary time evolution. Following our definition for discrete-time quantum walks, we suggest to measure each system in an ensemble only once, after different periods of time evolution. We discuss both random (Poisson process) and periodical measurements.

In section \ref{definition}. we give a definition for the P\'olya number,
and in section \ref{examples}. we use our definition with Poissonian
sampling for the periodical chain, and for the $d$-dimensional integer
lattices. In section \ref{periodical}. we compare our results with
periodical measurement for the systems that were considered in section
\ref{examples}. In section \ref{other}. we apply our methods that were used
in section \ref{examples}. for other graphs with known time evolution to
determin its P\'olya number.

\section{Definition of the P\'olya number}
\label{definition}

The continuous-time quantum walk was defined by Farhi and Gutmann \cite{farhigut}. We have a $G(V,E)$ undirected graph, and let us denote the vertices with $a=0\dots N-1$. Then corresponding to the graph we construct an $N$-dimensional Hilbert space with an orthogonal $\{ |a\rangle \}$ basis, where $a=0\dots N-1$ and $\langle a | b\rangle =\delta_{ab}$. The unitary time evolution is
\begin{gather}
\hat{U} (t)=e^{-i\hat{H}t},
\end{gather}
where the matrix elements of the Hamiltonian are
\begin{gather}
\langle a |\hat{H}| b \rangle =\left \{ \begin{array}{ll}
-\gamma, & \textrm{if } |a\rangle \textrm{ and } |b\rangle \textrm{ are neighbouring states,} \\
0, & \textrm{if } |a\rangle \textrm{ and } |b\rangle \textrm{ not neighbouring,} \\
k \gamma, & \textrm{if } |a\rangle=|b\rangle,
\end{array}
\right.
\label{hamiltoni}
\end{gather}
and $k$ is the degree of the vertex, and $\gamma$ is a time-independent constant. For simplicity, we can choose $\gamma =1$, it only rescales time. The wave function of the system after time $t$ is
\begin{gather}
|\Psi (t)\rangle=e^{-i\hat{H}t}|\Psi (0)\rangle,
\end{gather}
and the walk starts from the $|\Psi(0)\rangle=|0\rangle$ state.
We would like to calculate the probability that the walker ever returns to the starting point of the walk, {\it i. e.}, the P\'olya number. When defining such a quantity, we have to consider measurements on the system. At this point one has to deal with two problems. First, a measurement on a quantum system means an unavoidable disturbance, leading generally to a non unitary evolution.
Second, the time evolution is continuous unlike for the discrete-time quantum walks, thus the natural procedure for discrete-time walks, to make a measurement in each time-step cannot be applied here. 
The problem of the disturbance caused by a measurement can be avoided if we adopt the following procedure \cite{dkpolyaprl,dkpolyapra}:
each measurement is carried out on a different member from an ensemble of identically prepared and identically evolving systems, after a different time of the evolution. For the second problem of timing the measurements Varbanov {\it et al} \cite{varbanov} suggested the use of a Poisson process with rate $\lambda$ when defining the hitting time of continuous-time quantum walk. We will compare the random timing and the regular time steps for the measurement.

Let us define the P\'olya-number in the following way.
We measure the position of the walker with a series of von Neumann measurements. Every measurement is made on a new system from an ensemble containing identical systems that were evolved in the same way.
The probability that we measure the walker at the origin reads
\begin{gather}
\label{eq:p0}
p_0(t)=|\langle 0|e^{-i\hat{H}t}|0\rangle|^2\, ,
\end{gather}
for each system in the ensemble. The events corresponding to the measurements are independent, therefore the probability that the walker was not found at the origin until time $t_n$ can be expressed as
\begin{gather}
\bar{P_n}=\prod_{i=1}^n[1-p_0(t_i)] \, .
\end{gather}
The P\'olya-number can then be defined as the probability of the complementary event, by taking the limit $t \to \infty$
\begin{gather}
P=1-\prod_{i=1}^{\infty}[1-p_0(t_i)] \, .
\label{folytpolya} 
\end{gather}
It can be shown that the infinite product in the second term is zero, iff the following sum is divergent
\begin{gather}
\label{eq:sum}
\mathcal{S}=\sum_{i=1}^{\infty}p_0(t_i) \, ,
\end{gather}
and then the P\'olya number is 1 and the walker returns to the origin with certainty, otherwise there is a finite probability of escaping. 

The above definition provides a fixed value for the P\'olya number for a given sequence of
time instants $\{t_1,t_2,t_3,...\}$ where $t_1<t_2<t_3<\dots$. For example, this is the case for measurements following each other regularly with $t_{i+1}-t_i=T$. We will consider such periodical measurements in Section \ref{periodical}. 

When the timing of the measurements is chosen randomly, we can describe it with a continuous-time counting process and the P\'olya number becomes a random variable with respect to the realizations of the process.  The probability space $(\Omega,\mathcal{F},\mathbb{P})$ together with the counting process $\{X(t)|t\in \mathds{R} \}$ defines the time instants when a measurement occurs: $\{t_1,t_2,t_3,...\}=\omega\in \Omega$ are the elements of the event space. Thus, the 
P\'olya number should be considered as a random variable. Its expectation value can be expressed as
\begin{gather}
\nonumber
E[P]=1-\lim_{N\to \infty}  \int_{0}^{\infty}dT_1 \dots dT_N\\
 \, f(T_1,T_2,\dots T_N) \prod_{n=1}^N\left [1-
p_0\left (\sum_{i=1}^n T_i\right ) \right ] \, ,
\end{gather}
where $f(T_1,T_2,\dots T_N)$ is the joint probability of having the first jump in the counting process at $t_1=T_1$, the second at $t_2=T_1+T_2$, and so on. The counting process is a renewal process if the inter-measurement times $T_i$
are independent, identically distributed variables. In this case the expectation value of the P\'olya number reads
\begin{gather}
\nonumber
E[P]=1-\lim_{N\to \infty}  \int_{0}^{\infty}dT_1 \dots dT_N\\
 \, \prod_{i=1}^N f(T_i) \prod_{n=1}^N\left [1-
p_0\left (\sum_{i=1}^n T_i\right ) \right ]  \, .
\end{gather}

\section{Examples with Poissonian measurements}
\label{examples}

We apply our definition to calculate the P\'olya number for various examples. In this section we will use a Poisson process for sampling the times for the measurements. It means that the $\{t_i\}_{i=1}^{\infty}$ measurement-time sequence contains random variables having the property that the interarrival times $T_1=t_1, T_2=t_2-t_1, T_3=t_3-t_2, \dots $  are independent random variables, each with $f(T_i)=\lambda e^{-\lambda T_i}$ distribution, according to the definition of the one dimensional Poisson process with intensity $\lambda$ \cite{poisson}. In this case the expected value of the P\'olya number reads
\begin{gather}
\nonumber
E[P]=1-\lim_{N\to \infty} \lambda^N \int_{0}^{\infty}dT_1 \dots dT_N\\
 \,\exp\left (\sum_{k=1}^N-\lambda T_k\right )\prod_{n=1}^N\left [1-
p_0\left (\sum_{i=1}^n T_i\right ) \right ]  \, .
\label{egzakt}
\end{gather}
This expression is too difficult to calculate exactly, in our calculations we will use other methods to evaluate the P\'olya-number.

\subsection{Periodical chain}
\label{chain}

Let us consider a periodical chain that contains $N$ quantum states. We have to find the probability at the origin (\ref{eq:p0}), which is the square of the $(1,1)$ element of the $e^{-i\hat{H}t}$ matrix. It can be calculated easily if we recognize that the Hamiltonian is a symmetric cyclic matrix
\begin{gather}
\mathbf{C}(c_0,c_1,\hdots,c_{N-1})=\left (
\begin{array}{ccccc}
c_0 & c_1 & c_2 & \hdots & c_{N-1} \\
c_{N-1} & c_0 & c_1 & \hdots & c_{N-2} \\
c_{N-2} & c_{N-1} & c_0 &  \hdots &  c_{N-3} \\
\vdots & \vdots & \vdots & \ddots & \vdots \\
c_1 & c_2 & c_3 & \hdots & c_0
\end{array}
\right ) \, .
\end{gather}
A general matrix element of a matrix function of symmetric cyclic matrices can be explicitly expressed \cite{rozsapal}. If the matrix is in the odd class ($N=2n+1$ is odd), then
\begin{gather}
\{f(\mathbf{C})\}_{pq}=\frac{1}{2n+1}f \left ( c_0+2\sum_{\nu =1}^n c_{\nu} \right ) +\nonumber\\
+ \frac{2}{2n+1}\sum_{k=1}^n f \left ( c_0+ 2\sum_{\nu =1}^n c_{\nu} \cos \frac{2\nu k \pi}{2n+1}  \right ) \cos \frac{(p-q)2k\pi}{2n+1} \, ,
\end{gather}
 and if $N=2n$ is even, then
\begin{gather}
 \{f(\mathbf{C})\}_{pq}=\frac{1}{2n}\left \{ f\left ( c_0+c_{\nu} + 2\sum_{\nu =1}^{n-1} c_{\nu} \right )+ \right. \nonumber\\ \left.+ (-1)^{p+q} f \left ( c_0+(-1)^n c_n +  2\sum_{\nu =1}^{n-1} (-1)^{\nu} c_{\nu}   \right ) \right \} + \nonumber\\
+ \frac{1}{n} \sum_{k=1}^{n-1}f \left ( c_0 + (-1)^k c_n + 2  \sum_{\nu=1}^{n-1} \cos \frac{\nu k \pi}{n}\right ) \cos \frac{(p-q)k\pi}{n} \,  .
\end{gather}
From the definition of the Hamiltonian (\ref{hamiltoni}) the matrix elements for the periodical chain are 0 except for $c_0=-2i\gamma t$ and $c_1=c_{N-1}=i\gamma t$. Therefore, the probability that the walker is at the origin for odd $N$ reads
\begin{gather}
p_0(t)=\frac{1}{(2n+1)^2}+\frac{4}{(2n+1)^2}\sum_{k=1,j=0}^n\cos (2\gamma t \xi_{kj}) \, ,
\label{paratlan}
\end{gather}
where
\begin{gather}
\xi_{kj}=\cos \frac{2k\pi}{2n+1}-\cos \frac{2j\pi}{2n+1} \, ,
\end{gather}
and for even $N$:
\begin{gather}
p_0(t)=\frac{1}{2n^2}+\frac{1}{2n^2}\cos(4\gamma t)+\frac{1}{n^2}\sum _{j=0}^n\sum _{k=1}^{n-1} \cos (2\gamma  t\zeta_{kj})  \, ,
 \label{paros}
\end{gather}
where
\begin{gather}
\zeta_{kj}=\cos \frac{k\pi}{n} -\cos \frac{j\pi}{n} \, .
\end{gather}
In order to find the P\'olya number, we are interested in the limit of the sum (\ref{eq:sum}).
There exist time series for the measurement such that the sum is convergent. However, we will show that the probability of this event is zero. We could express the function $p_0(t)$ as a finite sum of cosine functions. Thus the extrema
of this analytic function cannot have an accumulation point other than infinity. On the other hand, it cannot tend to zero. Let us choose a small positive constant $\varepsilon$ ($0<\varepsilon<1$). We can estimate the sum by taking only values greater than $\varepsilon$
\begin{gather}
\sum_{i=1}^{\infty}p_0(t_i)\geq \sum_{p_0(t_i)>\varepsilon}p_0(t_i) \, .
\label{becsles}
\end{gather}
We are going to prove that the probability of having only a finite number of points greater than $\varepsilon$ is zero.
Let us denote the position of the $i$th minimum with $n_i$ and introduce $\delta_i$, the size of the interval around $n_i$ where $p(t)<\varepsilon$. Let the test period be $t=m_N$ and suppose that $M$ samples are taken from the function. We find that the probability of $j$ sample from the $M$ being greater than $\varepsilon$ reads
\begin{gather}
P(j)=\binom{M}{j}\left ( \frac{\sum_{i=1}^{N-1}\delta_i}{t} \right )^{M-j} \left ( \frac{t-\sum_{i=1}^{N-1}\delta_i}{t} \right ) ^j  \, .
\end{gather}
The second and the third factors can be rewritten as
\begin{gather}
P_\infty (j)=\lim_{M,t\to \infty} \binom{M}{j} \eta ^{M-j}\mu ^j =0 \, ,
\end{gather}
where $\eta<1$ (for a small enough $\varepsilon$ it cannot be 1, even in the limit of $t\to \infty$, because $p_0(t)$ is an almost periodic function) \rm and $\mu<1$ and does not depend on $M$.
We find that there are infinitely many points where $p_0 (t)$ is greater than  $\varepsilon$, therefore the sum
\begin{gather}
\mathcal{S}=\sum_{i=1}^{\infty}p_0(t_i)
\end{gather} 
is divergent. Thus we have proved that the walk on a periodical chain is recurrent, its P\'olya number is one. In the proof we did not use the $\lambda$ parameter, consequently the P\'olya number is independent of the parameter of the Poisson process.

 This proof can be applied for any finite graph. The probability that we measure the walker at the origin  for a finite system that contains $N$ quantum states  can written as a finite sum of cosine functions. This probability reads
\begin{gather}
 p_0(t)=\left | \sum_{n=0}^{N-1} \langle 0 | e^{-iE_nt} | q_n \rangle \langle q_n | 0 \rangle \right |^2 \, ,
\end{gather}
where $E_n$ are the eigenvalues and $|q_n\rangle$ are the orthonormalized eigenvectors of the Hamiltonian. From this formula we get
\begin{gather}
 p_0(t)=\sum_{n,m=0}^{N-1} Q_mQ_n\cos \left [(E_m-E_n)t \right ] \, ,
\end{gather}
where $Q_n=|\langle 0 | q_n \rangle |^2$ are reals and independent of $t$. Therefore for any finite system the P\'olya number equals one.

\subsection{One dimensional integer lattice}
\label{1dlattice}

To calculate the P\'olya number of the one dimensional lattice we need the probability of measuring the walker at the origin
\begin{gather}
p_0(t)=|\langle 0|e^{-i\hat{H}t}|0\rangle|^2 \, ,
\end{gather}
if we start the walker from the $|0\rangle$ state. The matrix elements of the evolution operator \cite{farhigut} read 
\begin{gather}
\langle k|e^{-i\hat{H}t}|j\rangle=i^{k-j}e^{-2it}J_{k-j}(2t) \, ,
\end{gather}
thus the probability at the origin is a Bessel function
\begin{gather}
p_0(t)=J_0^2(2t) \, .
\end{gather}
We have to examine the convergence of the	
\begin{gather}
 \mathcal{S}=\sum_{i=1}^{\infty}p_0(t_i)
\end{gather}
sum if we have a given $\left \{t_i \right \}_{i=1}^{\infty}$ measurement-time sequence. We can utilize the asymptotic form of the Bessel function
\begin{gather}
J_m(x)\sim \sqrt{\frac{2}{\pi x}}\cos\left ( x-\frac{m\pi}{2}-\frac{\pi}{4} \right ) \, .
\label{asszimptotikus}
\end{gather}
Since a finite number of terms does not change the convergence of a series, 
for the evaluation of the P\'olya number we have to consider the asymptotic convergence 
\begin{gather}
 \mathcal{S}_2=\sum_{i=1}^{\infty} \frac{\cos ^2\left ( 2t_i-\frac{\pi}{4} \right )}{t_i} \, .
\end{gather}
In the Appendix, we prove that the sum $\mathcal{S}_2$ with Poissonian sampling is divergent with probability 1.
Consequently, the sum
\begin{gather}
 \mathcal{S}=\sum_{i=1}^{\infty}p_0(t_i)
\end{gather}
is divergent with probability 1, therefore the continuous-time quantum walk on the line is recurrent, its P\'olya number equals 1.

\subsection{Higher dimensional integer lattices}
\label{dlattice}

In this section we consider the P\'olya number for a $d$-dimensional integer lattice. 
The probability factorizes to $d$-independent one-dimensional probabilities \cite{Mulken14,Aligrest}, thus the probability that we measure the walker at the origin reads
\begin{gather}
 p_0(t)=\left( \prod_{i=1}^d J_0(2t) \right)^2=\left( J_0(2t) \right)^{2d} \, .
\end{gather}
From the asymptotic form of the Bessel functions (\ref{asszimptotikus}), the envelope of $p_0(t)$ is $\dfrac{1}{t^d}$. We would like to determine the convergence of the sum
\begin{gather}
\mathcal{S}=\sum_{i=1}^{\infty}p_0(t_i) \, .
\end{gather}
If the above sum is convergent the walk is transient.
We can estimate the sum by
\begin{gather}
\mathcal{S}  \leq \sum_{t=1}^{\infty}g(t)\frac{1}{(t-1)^d} \, ,
\end{gather}
where $g(t)$ is the number of measurement points in the $[t-1;t]$ intervals, where $t\in\mathds{N}$ . Now, consider the probability that in each interval there are less measurement points than the corresponding $t$, it is the probability that $g(t)<t$. According to the definition of the Poisson distribution, the probability that there are less than $t$ measurement points in any unit interval reads
\begin{gather}
P\left(g(t)<t\right)=\sum_{k=0}^{t-1}\frac{\lambda^k}{k!}e^{-\lambda} \, .
\end{gather}
From this, the probability that our condition is fulfilled reads
\begin{eqnarray}
P\left(g(t)<t \, | \, \forall t\right)&=&\prod_{t=1}^{\infty}\left ( \sum_{k=0}^{t-1}\frac{\lambda^k}{k!}e^{-\lambda} \right ) \nonumber \\
&=&
\prod_{t=1}^{\infty}\left (1-  \sum_{k=t}^{\infty}\frac{\lambda^k}{k!}e^{-\lambda} \right ) \, .
\end{eqnarray}
The above infinite product tends to 0 if and only if the following double sum is divergent (see \cite{dkpolyapra}):
\begin{gather}
 \sum_{t=0}^{\infty}\sum_{k=t}^{\infty}\frac{\lambda^k}{k!}e^{-\lambda} =\infty \, .
\label{valsegkonverg}
\end{gather}
Since the summand of the first sum is decreasing faster than any power
\begin{gather}
\sum_{k=t}^{\infty}\frac{\lambda^k}{k!}e^{-\lambda}-\sum_{k=t+1}^{\infty}\frac{\lambda^k}{k!}e^{-\lambda}=\frac{\lambda^t}{t!}e^{-\lambda} \, ,
\end{gather}
thus the (\ref{valsegkonverg}) sum is convergent. Consequently, 
\begin{equation}
P\left(g(t)<t \, | \, \forall t\right) > 0 \, ,
\end{equation}
 i. e. the probability that in each interval there are less measurement points than $t$ is not 0. Thus, there is a nonzero probability that $g(t)<t$ and therefore with the same probability the sum $\mathcal{S}$ can be estimated as
\begin{gather}
\mathcal{S}<\sum_{t=1}^{\infty} t \frac{1}{(t-1)^d} \, ,
\end{gather}
which is convergent for $d\geq 3$. 

In summary, if the dimension of the lattice is 3 or greater then there is a finite probability that the walker never returns to the origin and thus the walk is transient.
A similar line of thought can be applied for dimension 2 and prove with a somewhat more lengthy but straightforward calculation that $P(g(t)<\sqrt{t})>0$ for the number of the measurement points in each interval. In this way one can prove the transience of the walk also for dimension 2.
 
One can estimate the expectation value of the P\'olya number from (\ref{egzakt}) using the first few factors from the product. Table \ref{table}. lists the approximate value of the P\'olya number for dimensions 2,3 and 4 by using the first three factors. In order to estimate the error we list the difference compared to the next order approximation.

\begin{center}
\begin{table}[ht]
 \begin{tabular}{|c|c|c|}

\hline
  $d$ & $E[P_3]$ & $E[P_4]-E[P_3]$ \\
\hline
2 &0.354 & 0.002\\
\hline
3 &0.2968 &0.0003 \\
\hline
4 &0.26374 & 0.00007\\
\hline

\end{tabular}
\caption{The expectation value of the P\'olya number for a regular integer lattice in 2, 3 and 4 dimensions with Poissonian timing of the measurements ($\lambda=1$). We have included three measurement points to estimate the expression (\ref{egzakt}). The contribution from the fourth measurement point is also evaluated, indicating the error of the estimation.}
\label{table}
\end{table}
\end{center}

\section{Examples with periodical measurements}
\label{periodical}

In this section we consider the case of measuring regularly, in $T$ time intervals in the same systems that we considered in the previous chapter. 

\subsection{Periodical chain}

For the periodical chain the probability is a finite sum of cosine functions, so it can be periodic or a non-periodic function, depending on the ratio of the coefficients in the arguments. If it is non-periodic, then with periodic measurements there will be an infinite number of points that are greater than a small constant, therefore the (\ref{eq:sum}) sum is divergent. If $p_0(t)$  is periodic and its minimum is zero (that is the case for example for a chain of length 4), and we measure with the same periodicity and timing of the first measurement, then the (\ref{folytpolya}) P\'olya-number is exactly 0. On the other hand, let us suppose that the timing of the measurements has some error. In this case the sum will become divergent and the P\'olya number is 1.

\subsection{Integer lattices}

The probability that we measure the walker at the origin for the one-dimensional lattice is $p_0(t)=J_0^2(2t)$.  One can see from the (\ref{asszimptotikus}) asymptotic form of the Bessel function that it is possible to asymptotically coincide with the zeros of the probability distribution if one measures with the right period. In this case the (\ref{eq:sum}) sum is convergent and the walk is transient, in contrast to the walk with Poissonian sampling, which is recurrent. If we allow for some small deviation $\delta$ in the timing of measurement, then we again find that the walk is recurrent with probability 1. 
The error of the measurement timing means that we take measurement points from $[nT-\delta;nT+\delta]$ intervals, where $T$ is the period of the measurement. The cosine function in the asymptotic form of the Bessel function is symmetric around the minimum point, therefore for the convergence of the 
\begin{gather}
 \mathcal{S}=\sum_{i=1}^{\infty}J_0^2(2t_i)
\end{gather}
sum we have to consider the convergence of the 
\begin{gather}
\sum_{n=1}^{\infty}\frac{\varepsilon_n}{nT+\delta}\geq\sum_{n:\Gamma\leq\varepsilon_n}\frac{\Gamma}{nT+\delta}
\label{eq:T}
\end{gather}
sum, where $0\leq\varepsilon_n\leq \varepsilon_{\delta}$ is a random variable, and $\cos^2[2(nT+\delta)-\frac{\pi}{4}]=\varepsilon_{\delta}$. For a given $0<\Gamma<\varepsilon_{\delta}$ constant the 
\begin{gather}
\sum_{n=1}^{\infty}\frac{\Gamma}{nT+\delta}=\sum_{n:\Gamma\leq\varepsilon_n}\frac{\Gamma}{nT+\delta}+\sum_{n:\Gamma>\varepsilon_n}\frac{\Gamma}{nT+\delta}
\label{gammasum}
\end{gather}
sum is divergent. If we choose  $\Gamma$  such that the probability of $\varepsilon_n$ being greater than $\Gamma$ is 0.5, then both sums on the right side in (\ref{gammasum}) are divergent and therefore (\ref{eq:T}) is divergent with probability one. Thus the walk becomes recurrent for the one-dimensional lattice if we allow any small error in the timing of the measurements. 

For the higher dimensional integer lattices the P\'olya number is less than one with the periodic measurement procedure, because the decay of the envelope of the $p_0(t)$ function is faster than $t^{-1}$. The actual value depends sensitively also on the timing of the first measurement.

\section{Other graphs}
\label{other}

In this section we show how we can apply our results and methods for systems where the asymptotic evolution of the probability at the origin is known.

For finite systems the $p_0(t)$ probability cannot tend to zero as explained in section \ref{chain}. A number of finite graphs have been discussed in the literature, for example the finite star graph \cite{Xu-star}, the complete graph, the cycle graph \cite{Jafar}, etc. The P\'olya number is one for all these finite graphs.

For an infinite graph, we have to consider the decay of the $p_0(t)$ probability. One can follow a similar analysis as in sections \ref{1dlattice} and \ref{dlattice}.  If  $p_0(t)$ has asymptotically the form $f(t)\cdot t^{-\alpha}$ with $\alpha\leq 1$ where $f(t)$ is a periodic or almost periodic analytical function, then with Poissonian sampling the walk is recurrent. If the  envelope decays faster ($\alpha > 1$), then the walk is transient. For example, the P\'olya-number of the honeycomb lattice is smaller than one, because for this graph $\alpha=2$ \cite{honey}. In a spidernet graph \cite{spidernet} $\alpha=3$, therefore the walk is transient. Finally, the probability that we measure the walker at the origin for an infinite Hermite graph is an exponential function \cite{Jafar}, therefore it converges to zero faster than any power, thus the walk is transient.

\section{Conclusions}
\label{conclusions}

The present definition for the P\'olya number of the continuous-time quantum walk is applicable for an arbitrary time-sequence of measurements. The measurements can be timed according to a stochastic process, with random waiting times. As a consequence, the P\'olya number, which is itself a probability due to the random nature of the measurement results on a quantum system, will become itself a random variable according to the probability space of the measurement time sequences.
We find that for some graphs one can determine its value with probability unity. In the general case it is a random variable depending on the timing of the measurements, thus one can use its expectation value.

We have focused on the recurrence properties of the walk for the periodical chain and  $d$-dimensional integer lattices using our definition with Poissonian sampling. The P\'olya number is determined by the decay of the probability that we measure the walker at the origin, similarly to the discrete-time quantum walk. 
For the periodical chain and the one-dimensional regular lattice the P\'olya number equals one, like for the classical as well as the discrete-time quantum walks. For higher dimensional integer lattices it is smaller than one, in contrast to the classical walk. Our proof for the periodical chain can be applied for any finite graph.

We have also discussed what happens if one measures periodically instead of the random Poissonian sampling. In this case a transient walk can be observed in systems where the P\'olya number with Poissonian sampling would be one. The apparent paradox is eliminated if the timing for the measurement is not ideal, then the P\'olya number is the same with Poissonian and periodically sampling.  When the decay of the probability at the origin is known for a given graph, the recurrence of the walk can be directly decided applying our definition. For transient walks the first few measurement points usually provides a good estimation for the P\'olya number.

\begin{acknowledgements}

The financial support by the Czech-Hungarian cooperation project (KONTAKT,CZ-11/2009) is gratefully acknowledged.

\end{acknowledgements}

\section*{Appendix}
\label{appendix}

In the Appendix we prove that
\begin{gather}
 \mathcal{S}=\sum_{i=1}^{\infty}p_0(t_i)
\end{gather}
where $p_0(t)=J_0^2(2t)$, is divergent with probability 1 for Poissonian sampling. The asymptotic form of the Bessel function reads
\begin{gather}
 J_0(x)\sim \sqrt{\frac{2}{\pi x}}\cos\left ( x-\frac{\pi}{4} \right ) \, ,
\end{gather}
thus the envelope of $p_0(t)$ is $\frac{1}{t}$. 

First we consider the 
\begin{gather}
\mathcal{S}\leq\mathcal{S}_1=\sum_{i=1}^{\infty}\frac{1}{t_i}
\end{gather}
sum.
We can use the following convergence criterion: let us suppose that the 
\begin{gather}
 \lim_{k\to \infty} \frac{a_k}{b_k}=\sigma \, 
\end{gather}
limit exists, where $0<\sigma<\infty$. Then $\sum_{k=1}^{\infty} a_k$ is convergent if and only if $\sum_{k=1}^{\infty} b_k$ convergent. We can choose
\begin{gather}
 a_k=\frac{1}{t_k} \, ,\quad b_k=\frac{1}{k}\, , \quad  \textrm{then} \quad \frac{a_k}{b_k}=\dfrac{k}{t_k}\, .
\end{gather}
The probability density function for the time of the $k$th measurement in the Poisson process reads
\begin{gather}
 F^{(k)}(t)=r^k \frac{t^{k-1}}{(k-1)!}e^{-rt}  \,.
\end{gather}
We know from the central limit theorem that this distribution converges to the normal distribution as $k\to \infty$:
\begin{gather}
 f^{(k)}(t)=\frac{\lambda}{\sqrt{2\pi k}} \exp \left ( -\dfrac{ \left ( t- \frac{k}{\lambda} \right ) ^2 \lambda ^2 }{2k} \right )\, .
\end{gather}
From this distribution we can calculate the probability, that $k$th measurement point, $t_k\in [A,B]$ :
\begin{gather}
 \mathds{P}=\int_A^B f^{(k)}(t) dt=\\
\nonumber
=\frac{1}{2} \left ( \textrm{erf} \left (\frac{k-A\lambda}{\sqrt{2k}} \right ) - \textrm{erf} \left (\frac{k-B\lambda}{\sqrt{2k}} \right ) \right ) \, ,
\end{gather}
where
\begin{gather}
 \textrm{erf}(x)=\frac{2}{\sqrt{\pi}} \int_0^x e^{-t^2} dt \, .
\end{gather}
We will see later that a good choice for the integration limits is
\begin{gather}
 A=\left ( 1-\frac{1}{\ln(k)} \right ) \frac{k}{\lambda} \, , \quad B=\left ( 1+\frac{1}{\ln(k)} \right ) \frac{k}{\lambda} \, .
\end{gather}
With this
\begin{gather}
 \mathds{P}=\textrm{erf} \left( \frac{\sqrt{k}}{\sqrt{2}\ln(k)} \right) \, .
\end{gather}
The limit of this probability from the definition of $\textrm{erf}(x)$ reads
\begin{gather}
 \lim_{k\to \infty} \mathds{P}=1 \, .
\end{gather}
Therefore, in the $k\to \infty$ case we find that 
\begin{gather}
 \left ( 1-\frac{1}{\ln(k)} \right ) \frac{1}{\lambda} \leq \frac{t_k}{k} \leq \left ( 1+\frac{1}{\ln(k)} \right ) \frac{1}{\lambda} \, ,
\end{gather}
with probability unity and if we take the limit on both sides
\begin{gather}
 \lim_{k\to \infty} \frac{t_k}{k}=\frac{1}{\lambda} \, .
\label{limit}
\end{gather}
From this result $\sigma=\lambda$ in the convergence criterion. We know that $\sum_{k=1}^{\infty} \frac{1}{k}$ is divergent, thus
\begin{gather}
 \mathcal{S}_1=\sum_{i=1}^{\infty} \frac{1}{t_i}
\end{gather}
is divergent with probability 1.

For our final result we have to consider the convergence of
\begin{gather}
 S_2=\sum_{i=1}^{\infty} \frac{\cos ^2\left ( 2t_i-\frac{\pi}{4} \right )}{t_i} \, .
\end{gather}
We know that
\begin{gather}
 \sum_{i=1}^{\infty} \frac{\cos ^2\left ( 2t_i-\frac{\pi}{4} \right )}{t_i}\geq \sum_{i\in P} \frac{\varepsilon}{t_i}=\mathcal{P} \, ,
\label{cos>P}
\end{gather}
where $0<\varepsilon\leq1$ is a small constant, and we introduced
\begin{gather}
  P=\{ i\in \mathds{N}: \cos^2(2t_i-\frac{\pi}{4})\geq\varepsilon \} \, , \\
\nonumber
Q=\{ i\in \mathds{N}: \cos^2(2t_i-\frac{\pi}{4})<\varepsilon \} \, .
\end{gather}
The following sum, which is divergent with probability unity, can be divided according to $P$ and $Q$
\begin{gather}
 \sum_{i=1}^{\infty} \frac{\varepsilon}{t_i}=\sum_{i\in P} \frac{\varepsilon}{t_i} + \sum_{i\in Q} \frac{\varepsilon}{t_i}=\infty \, .
\end{gather}
Consider the case if $\varepsilon=0.5$. Then the length of the $T$ period of the cosine function is divided into two equal parts with size $\frac{T}{2}$ and $t_i$ falls with equal probability in $P$ or $Q$. For large $i$ index, the $i$th element of the sum $\mathcal{P}$ will be near to the $2i$th element of the original sum. Therefore from (\ref{limit})
\begin{gather}
\lim_{k\to \infty} \frac{t_k^{(P)}}{k}=2\lim_{k\to \infty} \frac{t_k}{k}=\frac{2}{\lambda} \, ,
\end{gather}
 and we can apply the convergence criterion for $\mathcal{P}$. It is divergent, therefore from (\ref{cos>P}) $\mathcal{S}_2$ is also divergent with probability unity. The convergence of the $\mathcal{S}$ sum depends on the asymptotic behavior of the  Bessel function. Since $\mathcal{S}_2$ is divergent,  $\mathcal{S}$ will be divergent as well.

\end{document}